\documentclass[conference]{IEEEtran}
\IEEEoverridecommandlockouts
\pdfoutput=1

\usepackage[braket, qm]{qcircuit}
\usepackage{amsmath}
\usepackage{tikz}

\newcommand*\qcontrolcolor[2]{\push{\footnotesize \tikz[baseline=(char.base)]{
                              \node[shape=circle,draw,inner sep=0.6pt,color=#1] (char) {#2};}}
                              \qw}  

\newcommand*\onecontrol{\qcontrolcolor{black}{1}}
\newcommand*\twocontrol{\qcontrolcolor{black}{2}}
\newcommand*\threecontrol{\qcontrolcolor{black}{3}}

\usepackage{cite}
\usepackage{amsmath,amssymb,amsfonts} 
\usepackage{algorithmic}
\usepackage{graphicx}
\usepackage{textcomp}
\usepackage{xcolor}

\usepackage[normalem]{ulem}
\usepackage{amssymb}
\usepackage{amsmath}
\usepackage{bbm}
\usepackage[]{algorithm2e}
\usepackage{multirow}
\usepackage{graphicx}
\usepackage{mathtools}
\usepackage{listings}
\usepackage{booktabs}
\usepackage{xspace}
\usepackage{physics}
\usepackage{qcircuit}
\usepackage{fancyhdr}
\usepackage{subcaption}

\usepackage{rotating}
\usepackage{afterpage}
\usepackage{atbegshi}
\usepackage{floatpag}
\usepackage{zref-abspage,zref-user}
\usepackage{hyperref}
\usepackage{comment}

\usepackage{tikz}
\usepackage{pgfplots}
\usetikzlibrary{patterns}
\usetikzlibrary{arrows}
\usetikzlibrary{external}
\usetikzlibrary{shapes}
\usetikzlibrary{fadings,shapes.arrows,shadows}
\usetikzlibrary{shadows.blur}
\usetikzlibrary{shapes.symbols}
\usetikzlibrary{shapes.multipart}
\usetikzlibrary{positioning}

\newcommand\plotscale{0.8}  

\def\BibTeX{{\rm B\kern-.05em{\sc i\kern-.025em b}\kern-.08em
    T\kern-.1667em\lower.7ex\hbox{E}\kern-.125emX}}
\begin{document}

\title{Communication Trade Offs in \\ Intermediate Qudit Circuits\\
\thanks{This work is funded in part by EPiQC, an NSF Expedition
in Computing, under grants CCF-1730082/1730449; in part
by STAQ under grant NSF Phy-1818914; in part by NSF
Grant No. 2110860;by the US Department of Energy Office 
of Advanced Scientific Computing Research, Accelerated 
Research for Quantum Computing Program; and in part by 
NSF OMA-2016136 and the Q-NEXT DOE NQI Center.
FTC is Chief Scientist at Super.tech and an
advisor to Quantum Circuits, Inc.}
}

\fancypagestyle{firststyle}{%
  \fancyhf{}%
  \renewcommand{\headrulewidth}{0pt}
  \fancyfoot[C]{\scriptsize\vspace{-2em}%
    \copyright 2022 IEEE.  Personal use of this material is permitted.  Permission from IEEE must be obtained for all other uses, in any current or future media, including reprinting/republishing this material for advertising or promotional purposes, creating new collective works, for resale or redistribution to servers or lists, or reuse of any copyrighted component of this work in other works.%
  }%
}
\pagestyle{fancy}
\renewcommand{\headrulewidth}{0pt}
\fancyhf{}
\cfoot{\thepage}

\newcommand{\TODO}[1][TODO]{{\colorbox{blue}{\textbf{\textcolor{white}{#1}}}}}
\newcommand{\hide}[1]{}

\makeatletter
\newcommand{\rotateFigurePageForLabel}[1]{%
  \zlabel{#1}%
  \AtBeginShipout{%
  \ifnum\c@page=\zref@extractdefault{#1}{abspage}{0}
    \pdfpageattr{/Rotate 90}
  \fi}
}
\makeatother
\author{\IEEEauthorblockN{Andrew Litteken\textsuperscript{\textsection}}
\IEEEauthorblockA{\textit{Department of Computer Science} \\
\textit{University of Chicago}\\
litteken@uchicago.edu}
\and
\IEEEauthorblockN{Jonathan M. Baker\textsuperscript{\textsection}}
\IEEEauthorblockA{\textit{Department of Computer Science} \\
\textit{University of Chicago}\\
jmbaker@uchicago.edu}
\and
\IEEEauthorblockN{Frederic T. Chong}
\IEEEauthorblockA{\textit{Department of Computer Science} \\
\textit{University of Chicago}\\
chong@cs.uchicago.edu}}
\maketitle
\begingroup\renewcommand\thefootnote{\textsection}
\footnotetext{Equal contribution}
\endgroup

\thispagestyle{firststyle}

\begin{abstract}
Quantum computing promises speedup of classical algorithms in the long term. Current hardware is unable to support this goal and programs must be efficiently compiled to  use of the devices through reduction of qubits used, gate count and circuit duration.

Many quantum systems have access to higher levels, expanding the computational space for a device. We develop higher level qudit communication circuits, compilation pipelines, and circuits that take advantage of this extra space by temporarily pushing qudits into these higher levels.  We show how these methods are able to more efficiently use the device, and where they see diminishing returns.
\end{abstract}



\section{Introduction}

Quantum computing in the long-term promises speedup over classical algorithms for important algorithms like unordered search and integer factorization \cite{grovers}\cite{shor}. Unfortunately, we are many years away from a large-scale implementation of either of these algorithms because current quantum computers have limited number available hardware qubits \cite{ionq, ibmq, bristlecone, rigetti}, but large numbers are required to produce high quality error-correct logical qubits. In the near-term, however, there are many promising applications, like variational algorithms such as VQE and QAOA \cite{vqe}\cite{qaoa}, which aim to solve classical optimization problems using the quantum computer as a sort of accelerator in a classical-quantum optimization loop. For these types of algorithms, the size of the optimization problem we can effectively execute is bounded by device error rates, qubit coherence times, and ultimately by the number of physically available devices \cite{NISQ}.

In order to extend the boundary of what is currently computable, optimizations aim to minimize gate counts, circuit duration, or physical requirements. One such optimization has been the use of multivalued quantum logic. Many quantum systems naturally have access to infinitely many discreet levels beyond the usual binary qubit levels \cite{MS-gates}. At face value, rewriting quantum algorithms in a higher radix confers the same constant advantage in terms of gate requirements and expected circuit duration as it would classically, where advantage is due strictly due to space compression. In the long term, this approach may not be enormously advantageous as the space savings may not outweigh the increase in error, especially when the number of available devices increases significantly. In the near-term, space savings are critical.

Alternative approaches, such as the use of \textit{intermediate qudits} \cite{intermediate-qutrits}, rely on a mixed radix strategy where a smaller number of qudits are used, or they are used only for a short amount of time. While specialized in its use, the expected advantages are strong, for example in the generalized Toffoli decomposition (also the subject of this work) using qutrits (3 level quantum systems) temporarily enables a logarithmic depth (approximately circuit duration) decomposition with linear number of two-qudit gates and requires no additional ancilla space, scratch bits. The best known qubit-only circuits can obtain logarithmic depth with linear gate counts \cite{qubit-toffoli}. These circuits require linear amounts of additional space which makes efficient implementations infeasible on error-limited devices and bounding the program size on available hardware.

While the advantage conferred from clever circuit design is clear, these prior works omit two critical considerations from their constructions. First, these circuits are not compiled to any real device. While some devices promise all-to-all connectivity, most devices have limited connections between devices meaning programs must have movement operations inserted during compilation. This introduces large amounts of communication overhead, on average, which must be minimized \cite{swap-min}. Furthermore, swapping states different dimensional qudits requires tailored communication operations which get more expensive as the dimension of the inputs increases. Second, qubit gates and qudit gates are not created equal. From a circuit point of view, they take up one unit of time each, but when implemented on device this may not be the case. To execute gates on hardware, analog pulse sequences must be generated, for example via a process called optimal control. The duration of these pulse sequences determines the length of a given gates. In some cases, converging to a high quality pulse sequence of minimal length is difficult \cite{optimal-control-bspline, optimal-control-discrete}.

In this work, we consider a straight-forward compilation of intermediate qudit circuits to connectivity-limited superconducting devices. To do so, we introduce swap gates designed explicitly to communicate qudits of different dimensions, for example qubits with ququarts (4 level systems). Second, we consider realistic gate times obtained via optimal control for superconducting based quantum architectures to better study expected circuit durations beyond circuit depth. Together, we use these compilation tools to study one representative implementation of the generalized Toffoli using different levels of intermediate qudits. 

Our contributions are the following:
\begin{itemize}
    \item Detail efficient decompositions of SWAP gates between qudits of different dimensions
    \item A simple compilation pipeline transforming intermediate qudit circuits into ones which obey connectivity-limited hardware constraints
    \item Introduce intermediate ququart and ququint (5 level system) implementations of the generalized Toffoli, a critical circuit component for many larger algorithms
    \item Evaluation of the near-term benefits conferred by using intermediate qudits using realistic gate times for superconducting devices.
\end{itemize}

\section{Background}
\label{sec:background}
\begin{table}
\centering
\begin{tabular}{lll}
\textbf{Qudit Levels} & \textbf{Interaction Time (ns)} & \textbf{Swap Time (ns)} \\ \hline \\
1 & 30 & - \\
2 & 50 & - \\
3 & 50 & - \\
4 & 50 & - \\
0, 1* & 150 & 600 \\
0, 2 & 500 & 1200 \\
0, 3 & 500 & 1500 \\
0, 4 & 600 & 1800 \\
1, 1 & 500 & 900 \\
1, 2 & 500 & 1200 \\
1, 3 & 500 & 1500 \\
1, $4^*$ & 600 & 1800 \\
2, $2^*$ & 675 & 2950 \\
2, 3 & 850 & 5000 \\
2, $4^*$ & 1025 & 7050 \\
3, 3 & 850 & 5000 \\
3, $4^*$ & 1025 & 7050 \\
4, $4^*$ & 1200 & 7500 \\
\end{tabular}
\caption{\label{tab:times} Times used for various gates across different levels of qudits. An asterisk indicates an interpolated value.}
\vspace{-1.0em}
\end{table}
Classically, the basic computation unit is the bit which takes the value of either 0 or 1. In the quantum setting, we often consider the quantum bit (qubit) the most fundamental unit. For many quantum technologies, such as superconducting qubits and trapped ions, the implementation naturally has access to infinitely many discrete levels which can be truncated to any dimension $d$ giving qu\textit{dits} which exist as linear superpositions superpositions of $d$ levels as $\ket{\psi} = \alpha_0\ket{0} + \alpha_1\ket{1} + ... + \alpha_{d-1}\ket{d-1}$ where if we choose $d = 2$ we recover the qubit. 

For a variety of reasons, such as increasing number of error channels which become harder to control, using large numbers of states is often impractical and instead we should carefully choose the computing radix based on our target applications and available hardware. For example, measurement (the process of collapsing a quantum state to a classical value) of high level systems is often challenging for trapped ions and if possible we should try to measure only qubits.

To manipulate quantum states we apply \textit{gates}, can be represented as unitary matrices. For the most part, hardware supports at most gates on 1 or 2 inputs and all larger gates must be synthesized directly from smaller ones. To obtain an answer from the device, the quantum state must be measured, a non-unitary gate which collapses the state to any of the basis elements. For this work, we consider a basis set which consists of the generalized versions of the $X_{+k}$ gates and the $X_{i, j}$ gates which are classical permutations of the basis elements. The first behaves by shifting every basis element's coefficient $+k$ modulo the dimension of the system. The second behaves by swapping the coefficients of the $i$-th and $j$-th states and leaving all other coefficients the same \cite{MS-gates}. We also consider the controlled versions of these gates. In some decompositions we will use Toffoli-like gates on higher dimensions, which have constant depth decompositions into 1 and 2 qudit gates.

Current quantum hardware also only supports interactions between some pairs of devices, usually indicated by the hardware \textit{connectivity graph} which for many technologies is usually sparse. Qudit states can be moved around the device with special communication operations, often SWAP gates, in order to manipulate arbitrary pairs of qudits. The number of communication operations depends both on the connectivity of the input program and the connectivity of the underlying hardware.

In order to execute programs on a given hardware target, input programs must be compiled. This usually amounts to several key steps: circuit optimization and decomposition, mapping \cite{mapping}, routing \cite{swap-min}, and scheduling \cite{scheduling}. All input circuits must be decomposed into the hardware's basis gate set which is usually kept small in order to reduce calibration overheads. Efficient decompositions are important to keep gate counts, depth, and space low. Some decompositions require the use of additional qudits called \textit{ancilla} to be efficient which are scratch bits which begin and end in a known state.

Quantum hardware is error-prone. Gates can fail and qudit states decay over time, so each of these steps is critical to program success - we must minimize total operations and total execution time. Prior work in quantum multivalued logic has focused primarily on gains obtained from the first of these steps: circuit optimization. Significant gate count and depth advantages can be obtained using \textit{intermediate qudits} which means that inputs and outputs of an input quantum program are binary, but are allowed to temporarily access higher level states during computation. The key observation for the advantage is \textit{localizing} the additional space which would normally take the form of ancilla \cite{intermediate-qutrits}. For the near- to intermediate- term, space is severely restricted meaning ancilla counts should be kept low to maximize the amount of computational space.

While gate counts and circuit depth (the length of the circuit's critical path) are often good indicators of a circuit's execution time, it is important to consider the physical realization of such gates on hardware. Via a process called \textit{optimal control} analog pulse sequences can be produced for a given unitary. Typically, durations are chosen before hand and the goal is to produce a high fidelity (quality) implementation of the gate. Finding optimal pulse durations is challenging, and is expected to scale quadratically with the dimension of the input unitary. We consider gate times obtained via optimal control to accurately account for communication time costs and total circuit duration. While hardware supports a limited gate set, optimal control procedures can permit us to synthesize any unitary. But, we limit this to a small gate set on bounded numbers of qudits to limit classical overhead. Instead, we can interpolate to predict other gate durations or produce pulses for gates in the decomposition. For space reasons we have omitted a full discussion of optimal control, see \cite{optimal-control-bspline, optimal-control-discrete} for more information. Gate times for this work are listed in Table \ref{tab:times}.  The times for qubit-qubit interactions and swaps, and single qubit interactions are known from gate times from devices such as IBM's hardware.  Times for qubit-quqart, quqart-ququart, and single ququart interactions and swaps were found via optimal control. Gate durations for qutrits and ququints were found via a linear interpolation between these points.  The exact times for any gate also depends on the exact Hamiltonian used to model the underlying device or other physical restrictions. The numbers obtained here are from a standard superconducting Hamiltonian, as would be found for IBM's hardware \cite{ibmq}.

As seen in table \ref{tab:times}, as the radix increases, so does the time required to perform an operation on each qudit.  As the highest energy level increases, the pulses required to achieved the desired result becomes more complex as it needs to satisfy more constraints and transitions, elongating the necessary pulse duration.  There are several examples of how this could be achieved physically \cite{guide-superconducting-qudits}.  These increased radix devices also have higher rates of decoherence, but if the circuit depth and communication this will be balanced by a shorter circuit duration.  While this work does not focus on other architectures, the presented circuits would be valid, but would a new set of control experiments \cite{trapped-ion-qutrits}.  However, we would expect similar scaling.

\section{Communication Circuits}

\begin{figure}
\centering

  \centering
  \makebox[.2\linewidth][c]{
    \input{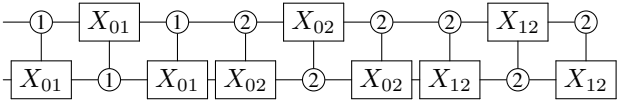}
  }
  \captionof{figure}{The decomposition of a qutrit based SWAP gate.}
  \label{fig:qutrit-swap-decomp}
\vspace{-1.0em}
\end{figure}

Current hardware has limited connectivity and only qudits which are adjacent on hardware may interact. Communication operations, called SWAPs, are required to move qudit states around the device. Here, we present a generalized decomposition of the SWAP gate taking in qudits of any dimension.

The decomposition of the qudit SWAP gate is straightforward and follows from the qubit-qubit swap. The key idea is to consider size 2 subsets of the basis elements. For qubits, this amounts to the subset $\{0, 1\}$ for which the decomposition is three 1 controlled $X_{01}$ flips. For qutrit SWAPs we now have three possible subsets $\{0, 1\}$, $\{0, 2\}$ and $\{1, 2\}$ and then perform partial SWAPs as if the qudits exist only in the subspace spanned by the elements of the subset. Here, we would do three 1 controlled $X_{01}$ flips followed by three 2 controlled $X_{02}$ flips and then three 2 controlled $X_{12}$ flips. The control value can be either of the elements of the subset.  

From a permutation point of view, a SWAP is decomposed into $O(n^2)$ 2-cycles where $n$ is the dimension of both qudits. For communication between qudits of different dimensions the same strategy applies but the scaling is $O(nm)$ where $n,m$ are the dimensions of the two input qudits. There have been other approaches for generalized SWAP gates by using more contrived basis elements \cite{qutrit-swaps, contrived-qudit-swaps}, but we find these to be intuitive and match the expected asymptotic scaling that optimal control predicts for gate durations. Additionally, this version of the SWAP gate will handle mixed radix inputs. That is, if one device is in a different radix from the other, the same set of gates for the higher radix SWAP can be used as would be used if they were both higher radix devices. In Figure \ref{fig:qutrit-swap-decomp} we show a swap gate decomposition for two qutrits.
\label{sec:communication}

\section{Compilation}

\begin{figure}
\centering
\begin{minipage}{.2\textwidth}
  \centering
  \makebox[.2\linewidth][c]{
    \input{figs/qubit-cnu.qcircuit}
  }
  \captionof{figure}{A logarithmic depth Generalized Toffoli circuit using qubits with 5 controls and 4 ancilla.}
  \label{fig:qubit-cnu}
\end{minipage}
\quad\quad
\begin{minipage}{.2\textwidth}
  \centering
  \makebox[.2\linewidth][c]{
    \input{figs/qutrit-cnu.qcircuit}
  }
  \captionof{figure}{A logarithmic depth Generalized Toffoli Circuit using qutrits with 7 controls.}
  \label{fig:qutrit-cnu}
\end{minipage}
\vspace{-1.0em}
\end{figure}

\begin{figure*}
\centering
\begin{minipage}{.45\textwidth}
  \centering
  \makebox[.4\linewidth][c]{
    \input{figs/ququart-cnu.qcircuit}
  }
  \captionof{figure}{A logarithmic depth Generalized Toffoli Circuit using ququarts with 7 controls.}
  \label{fig:ququart-cnu}
\end{minipage}
\quad\quad
\begin{minipage}{.45\textwidth}
  \centering
  \makebox[.4\linewidth][c]{
    \input{figs/ququint-cnu.qcircuit}
  }
  \captionof{figure}{A logarithmic depth Generalized Toffoli Circuit using ququints with 10 controls.}
  \label{fig:ququint-cnu}
\end{minipage}
\vspace{-1.0em}
\end{figure*}

Since many qudits will be in an higher level state, we must use qudit SWAP gates to effectively route the circuit on a device. Determining a favorable mapping and efficient routing scheme is important to effectively utilizing a quantum architecture when executing a quantum circuit \cite{mapping}\cite{scheduling}.  Placing or moving qudits that interact often far apart from one another on an architecture will require extra swap gates to move qudits within range of one another.  We adapt previously developed methods used for traditional qubit mapping and routing to effectively schedule qudit operations on a high radix device.  The methods described here are very similar to previously developed methods \cite{map3, map1, routing1, routing2}, but have the extra constraint of attempting to reduce time by utilizing as few high-radix communication operations as possible rather than reducing communication overall.  An implementation of this compiler can be found here \cite{github-link}.

We first decompose any gates that are not native to the qudit architecture.  For example, the Toffoli gate is a three qubit gate that can not be executed natively, and is decomposed to 6 CNOT gates, and 14 one qubit gates. Similarly, we decompose the qudit Toffoli gate as well according to \cite{di2011elementary}.  Then, we attempt to place the qudits that need to interact often in clusters to prevent extra communication costs based on the decomposed circuit.  Similar to other mapping methods, we first find the interaction weights between each qudit $w(u, v)$ interaction using:

\vspace{-1.0em}
\begin{align*}
    w(u, v) = \sum_{o \in ops} \mathbbm{1}(u, v \in o.qubits)
\vspace{-1.0em}
\end{align*}

\begin{figure*}
  \centering
  \begin{minipage}{.45\textwidth}
    \centering
    \scalebox{\plotscale}{
    \makebox[.4\linewidth][c]{
%
%
%
%
\begin{tikzpicture}[baseline,scale=1,trim axis left,trim axis right]
\pgfplotsset{every tick label/.append style={font=\small}}
\pgfplotsset{every axis label/.append style={font=\small}}

    \begin{axis}[
        name=plot3,
        title={Depth of Circuit, Before Routing},
        xlabel={Number of Controls},
        ylabel={Depth of Circuit},
        width={\columnwidth},
        height={0.8*\columnwidth},
        xmin=0, xmax=69, ymin=0, ymax=174.46,
        ,
        legend style={
            draw=none,
            at={(1.5,1)},
            anchor=north,
            font=\small},
        ,
        clip=false,
        axis line style={draw=none},
        tick style={draw=none},,
        legend columns=4, legend cell align={left}, legend style={at={(1.4,1.02)},anchor=north}, mark size=3pt,
    ]
        \addplot[color={rgb,255:red,255;green,113;blue,0}, thick, mark=x, ]
            table[x=number-of-controls 0, y=log-qubit depth-of-circuit 0, col sep=comma]
            {data/pre-depth-depth-of-circuit-before-routing.csv}
        ;
        \addlegendentry{Log Qubit};

        \addplot[color={rgb,255:red,209;green,216;blue,0}, thick, mark=+, ]
            table[x=number-of-controls 1, y=log-qutrit depth-of-circuit 1, col sep=comma]
            {data/pre-depth-depth-of-circuit-before-routing.csv}
        ;
        \addlegendentry{Log Qutrit};

        \addplot[color={rgb,255:red,0;green,205;blue,110}, thick, mark=o, mark size=2pt, ]
            table[x=number-of-controls 2, y=log-ququart depth-of-circuit 2, col sep=comma]
            {data/pre-depth-depth-of-circuit-before-routing.csv}
        ;
        \addlegendentry{Log Ququart};

        \addplot[color={rgb,255:red,125;green,0;blue,255}, thick, mark=square, ]
            table[x=number-of-controls 3, y=log-ququint depth-of-circuit 3, col sep=comma]
            {data/pre-depth-depth-of-circuit-before-routing.csv}
        ;
        \addlegendentry{Log Ququint};

    \end{axis}

\end{tikzpicture}
    }
    }
    \captionof{figure}{The depth of the Generalized Toffoli gates at different levels of qudits before communication gates are added to the circuit.}
    \label{fig:pre-depth}
  \end{minipage}
  \quad\quad
  \begin{minipage}{.45\textwidth}
    \centering
    \scalebox{\plotscale}{
    \makebox[.4\linewidth][c]{
%
%
%
%
\begin{tikzpicture}[baseline,scale=1,trim axis left,trim axis right]
\pgfplotsset{every tick label/.append style={font=\small}}
\pgfplotsset{every axis label/.append style={font=\small}}

    \begin{axis}[
        name=plot3,
        title={Depth of Circuit, After Routing},
        xlabel={Number of Controls},
        width={\columnwidth},
        height={0.8*\columnwidth},
        xmin=0, xmax=69, ymin=0, ymax=2278.96,
        ,
        legend style={
            draw=none,
            at={(0.5,1)},
            anchor=north,
            font=\small},
        ,
        clip=false,
        axis line style={draw=none},
        tick style={draw=none},,
        legend columns=2, legend cell align={left}, legend style={at={(0.5,1.02)},anchor=north}, mark size=3pt,
    ]
        \addplot[color={rgb,255:red,255;green,113;blue,0}, thick, mark=x, ]
            table[x=number-of-controls 0, y=log-qubit depth-of-circuit 0, col sep=comma]
            {data/pre-depth-depth-of-circuit-after-routing.csv}
        ;

        \addplot[color={rgb,255:red,209;green,216;blue,0}, thick, mark=+, ]
            table[x=number-of-controls 1, y=log-qutrit depth-of-circuit 1, col sep=comma]
            {data/pre-depth-depth-of-circuit-after-routing.csv}
        ;

        \addplot[color={rgb,255:red,0;green,205;blue,110}, thick, mark=o, mark size=2pt, ]
            table[x=number-of-controls 2, y=log-ququart depth-of-circuit 2, col sep=comma]
            {data/pre-depth-depth-of-circuit-after-routing.csv}
        ;

        \addplot[color={rgb,255:red,125;green,0;blue,255}, thick, mark=square, ]
            table[x=number-of-controls 3, y=log-ququint depth-of-circuit 3, col sep=comma]
            {data/pre-depth-depth-of-circuit-after-routing.csv}
        ;

    %
    \end{axis}

\end{tikzpicture}
    }
    }
    \captionof{figure}{The depth of the Generalized Toffoli gates at different levels of qudits after communication gates are added to the circuit.}
    \label{fig:post-depth}
  \end{minipage}
  \vspace{-1.0em}
  \end{figure*}

The most used qudit is placed in the center qudit of the quantum architecture.  From this point, we select the qudit that maximizes the sum of weights to the already placed qubits. The selected qudit is placed in an adjacent qudit to already placed qudits according to which qudit minimizes the sum of the products of time to interaction with a qudit from the location by the interaction weight to the qudit:

\vspace{-1.0em}
\begin{align*}
    m(u) = \sum_{v \in P} w(u, v) \times d(\varphi(u), \varphi(v))
\vspace{-1.0em}
\end{align*}

where $\varphi$ is the mapping from virtual to physical qudits and $P$ is the already placed qudits.  This keeps qudits that interact often close to one another, giving the scheduling a well placed starting point to begin routing the operations in the circuit.

To route the qudit operations, we attempt to disrupt the placement of the qudits will that interact frequently as little as possible.  By minimizing disruption, rather than naively moving qudits along the shortest path, we can potentially avoid extra swaps required bring these qudits back within range of one another.  We have the following scoring function representing the state of the mapped qudits at a given time in circuit execution and select the best potential qudit to swap with:

\vspace{-1.0em}
\begin{align*}
    s(Q, w, d) = \sum_{u, v \in Q \times Q} w(u, v) \times d(\varphi(u), \varphi(v))
\vspace{-1.0em}
\end{align*}

Where $Q$ represents the qudits in the circuit, $w$ is the interaction weights of the remaining operations to execute, and $d$ is the time, including swaps for two physical qudits to interact.  We find the difference between the current score, and the new mapping resulting from a SWAP.  We perform whichever SWAP minimizes the difference between the current score and the new score. Once an operation is completed, we remove 1 from the weight of the interacting pair as it is no longer relevant to the placement of the current qudits.

\label{sec:compilation}

\section{Benchmarks}

We focus on the Generalized Toffoli, or N-Controlled X gate as our benchmark.  This circuit is easily generalized to higher level quantum systems with similar constructions at each level. At the qubit level, this gate can be implemented with any number of ancilla qubits \cite{qubit-toffoli}.  But, to achieve this circuit in logarithmic depth we need to use $n-2$ ancilla for $n$ number of controls, significantly reducing the amount of usable space available to execute a quantum circuit.  An example of this circuit for 5 controls with 3 ancilla qubits is in Figure \ref{fig:qubit-cnu}.

An ancilla-free log depth version of the generalized Toffoli has also been developed for qutrit capable quantum systems \cite{intermediate-qutrits}. By using intermediate qutrits, this circuit uses 1 and 2 controlled +1 gates to reduce the number of qubits used and number of gates used.  The set of controls are treated as nodes in a tree, recursively stepping down the child nodes in the tree to create the circuit.  The child nodes target their parent node with a controlled qudit Toffoli gate which could increment or decrease the state of a qubit rather than performing an X gate. In the case of the leaf nodes, we use a 1 controlled qudit Toffoli gates.  Then, any internal nodes use a 2 control since their states will have been increased from 1 to 2 if the leaf qubits were in the 1 state.  Finally, the qubit representing the root of the tree performs a 2 controlled X gate on the target qubit.  We then reverse the controlled qudit Toffoli gates to return the input qutrit to their original, non-elevated state.  An example of this circuit with 7 controls is seen in Figure \ref{fig:qutrit-cnu}.

We expand on this idea by constructing generalized Toffoli circuits that take advantage of even higher level quantum systems.  We use the same tree-like framework developed for qutrit circuits, but can use the additional computational space afforded to us by these systems. Specifically, we use this extra space to remove the use of the gates involving three qudits, replacing them with two gates that use only two qudits each.  While this may seem like an increase in gate count, the three count qudit gates is decomposed into many more one and two qudit gates.  For qudit Toffoli gates, 6 two qudit gates and 8 single qudit gates, which is significantly higher.  In the ququart case, we use each leaf control to target its parent qudit with a +1 gate.  If the controls are in the 1 state, the target will end in the 3 state after being targeted by two controlled +1 gates.  Then the internal controls are 3 controlled +1 gates.  Similarly, the final controlled X gate becomes a 3 controlled X gate.  An example of this can be seen in Figure \ref{fig:ququart-cnu}.

A similar strategy can be used for any $n$ level system.  We split create the control tree into a tree with $n-2$ children rather than the original 2 children in the qutrit and ququart case. An example of this expansion into a generalized Toffoli gate for 5 level qudits (ququints) can be seen in Figure \ref{fig:ququint-cnu}.

\label{sec:benchmarks}

\section{Results and Discussion}

In this work we observe the changes in gate count, depth, duration, and space time product of a given circuit as we adjust our benchmark to utilize the higher radix levels of a quantum system, up to ququint level devices for the generalized Toffoli gate.  Both gate count and depth of a circuit on their own are helpful metrics to compare the runtime of two different quantum circuits.  However, as mentioned, different gates have different lengths, and may use more qubits through ancilla. So, it is necessary to examine the duration of the circuit as well.  The product of the duration and number of qubits is also used as the space-time product.  Minimizing this metric indicates an increase efficiency of the compiled circuit on a given device.

When measuring the time from these circuits, we construct a directed graph with nodes representing the operations, and edges representing dependencies based on qudits used.  We are able to label these edges with the time it will take based on the current levels of the qudits at that time.  This allows us to find the longest path through the graph, giving us the duration of the critical path, and the duration of the overall circuit.  We examine each circuit on a 12 qudit by 12 qudit grid architecture.  This provides a middle ground between the structure current quantum architectures, and devices with higher connectivity.  It also is a large enough architecture to examine the scalability of a circuit as they grow in size.

\begin{figure*}
  \centering
  \begin{minipage}{.45\textwidth}
    \centering
    \scalebox{\plotscale}{
    \makebox[.35\linewidth][c]{
%
%
%
%
\begin{tikzpicture}[baseline,scale=1,trim axis left,trim axis right]
\pgfplotsset{every tick label/.append style={font=\small}}
\pgfplotsset{every axis label/.append style={font=\small}}

    \begin{axis}[
        name=plot3,
        title={Circuit Duration},
        xlabel={Number of Controls},
        ylabel={Circuit Duration (ns)},
        width={\columnwidth},
        height={0.8*\columnwidth},
        xmin=0, xmax=69, ymin=0, ymax=184590.0,
        ,
        legend style={
            draw=none,
            at={(0.5,1)},
            anchor=north,
            font=\small},
        ,
        clip=false,
        axis line style={draw=none},
        tick style={draw=none},,
        legend columns=4, legend cell align={left}, legend style={at={(1.4,1.02)},anchor=north}, mark size=3pt,
    ]
        \addplot[color={rgb,255:red,255;green,113;blue,0}, thick, mark=x, ]
            table[x=number-of-controls 0, y=log-qubit circuit-duration-s 0, col sep=comma]
            {data/circuit-duration-circuit-duration.csv}
        ;
        \addlegendentry{Log Qubit};

        \addplot[color={rgb,255:red,209;green,216;blue,0}, thick, mark=+, ]
            table[x=number-of-controls 1, y=log-qutrit circuit-duration-s 1, col sep=comma]
            {data/circuit-duration-circuit-duration.csv}
        ;
        \addlegendentry{Log Qutrit};

        \addplot[color={rgb,255:red,0;green,205;blue,110}, thick, mark=o, mark size=2pt, ]
            table[x=number-of-controls 2, y=log-ququart circuit-duration-s 2, col sep=comma]
            {data/circuit-duration-circuit-duration.csv}
        ;
        \addlegendentry{Log Ququart};

        \addplot[color={rgb,255:red,125;green,0;blue,255}, thick, mark=square, ]
            table[x=number-of-controls 3, y=log-ququint circuit-duration-s 3, col sep=comma]
            {data/circuit-duration-circuit-duration.csv}
        ;
        \addlegendentry{Log Ququint};

    \end{axis}

\end{tikzpicture}
    }
    }
    \captionof{figure}{The circuit duration of the Generalized Toffoli at different controls for different qudit levels.}
    \label{fig:circuit-duration}
  \end{minipage}
  \quad\quad
  \begin{minipage}{.45\textwidth}
    \centering
    \scalebox{\plotscale}{
    \makebox[.35\linewidth][c]{
%
%
%
%
\begin{tikzpicture}[baseline,scale=1,trim axis left,trim axis right]
\pgfplotsset{every tick label/.append style={font=\small}}
\pgfplotsset{every axis label/.append style={font=\small}}

    \begin{axis}[
        name=plot3,
        title={Space Time Product},
        xlabel={Number of Controls},
        ylabel={Space Time Product},
        width={\columnwidth},
        height={0.8*\columnwidth},
        xmin=0, xmax=69, ymin=0, ymax=18369351.8,
        ,
        legend style={
            draw=none,
            at={(0.5,1)},
            anchor=north,
            font=\small},
        ,
        clip=false,
        axis line style={draw=none},
        tick style={draw=none},,
        legend columns=2, legend cell align={left}, legend style={at={(0.5,1.02)},anchor=north}, mark size=3pt,
    ]
        \addplot[color={rgb,255:red,255;green,113;blue,0}, thick, mark=x, ]
            table[x=number-of-controls 0, y=log-qubit space-time-product 0, col sep=comma]
            {data/space-time-product-space-time-product.csv}
        ;

        \addplot[color={rgb,255:red,209;green,216;blue,0}, thick, mark=+, ]
            table[x=number-of-controls 1, y=log-qutrit space-time-product 1, col sep=comma]
            {data/space-time-product-space-time-product.csv}
        ;

        \addplot[color={rgb,255:red,0;green,205;blue,110}, thick, mark=o, mark size=2pt, ]
            table[x=number-of-controls 2, y=log-ququart space-time-product 2, col sep=comma]
            {data/space-time-product-space-time-product.csv}
        ;

        \addplot[color={rgb,255:red,125;green,0;blue,255}, thick, mark=square, ]
            table[x=number-of-controls 3, y=log-ququint space-time-product 3, col sep=comma]
            {data/space-time-product-space-time-product.csv}
        ;

    %
    \end{axis}

\end{tikzpicture}
    }
    }
    \captionof{figure}{The space time product, or number of qudits used by duration, at different controls for different qudit levels.}
    \label{fig:space-time-product}
  \end{minipage}
  \vspace{-1.0em}
  \end{figure*}

\subsection{Benefits of Qutrits}
Previous examinations of the intermediate qutrit generalized Toffoli gate did not examine how it may be affected by routing the circuit on an limited connectivity architecture.  In Figure \ref{fig:pre-depth}, we see that the depth of the log-depth qutrit generalized Toffoli gate is greater than the depth of the qubit generalized Toffoli gate, despite the fact that it requires more Toffoli gates to implement the qubit gate.  Once SWAP gates are integrated into the circuit via the qudit compilation framework, in Figure \ref{fig:post-depth}, the depth of the qutrit generalized Toffoli gate is significantly lower than the qubit generalized Toffoli gate.  However, referring to Table \ref{tab:times}, the higher the level of the operations, the longer the times.  Upon measuring the duration of a circuit after inserting SWAPs, seen in Figure \ref{fig:circuit-duration} we see that at smaller numbers of controls, qutrit circuit duration is much lower than qubit circuit duration, but as the number of controls increases, these values reverse.  Where the qutrit circuit was half the duration of the qubit circuit from 5 to 15 controls, they are 1.5 times the duration from 35 controls onwards.  The extra time required by the SWAP gates exceeds the benefits gained from using fewer gates.

However, the construction of the qutrit circuit does not require any ancilla.  To take this into account, we examine the change in space time product of the qutrit versus qubit generalized Toffoli gates in Figure \ref{fig:space-time-product}. As the qutrit circuit does not require extra qubits, it can achieve a much lower space time product.  This is especially evident when the qutrit circuit has a low duration, starting at a ratio of 3 to 1 from qubit to qutrit space time ratio.  As the number of controls increases, this ratio converges to 1.7 to 1 space time ratio from qubits to qutrits, which matches the expected constant increase in computational space from qubits to qutrits, $\text{log}_2(3) = 1.58$.

\subsection{Benefits of Ququarts over Qutrits}

Moving from the qutrit circuit to the ququart circuit does not reduce the number of the qudits.  However, since we do not need to decompose any three qudit gates, the circuit requires fewer gates overall.  The difference in depth before and after inserting SWAPs can be seen in Figure \ref{fig:pre-depth} and Figure \ref{fig:post-depth}.  After routing, the depth of the qutrit circuit is 5 to 6 times greater than the ququart circuit, following from the ququart circuit using two two qudit gates, with a depth of two, rather than a Toffoli gate, which uses eight single qudit gates and six two qudit gates and has a depth of 11.

Taking the duration of these ququart gates into account in Figure \ref{fig:circuit-duration}, the reduction in gate count and depth generally outweighs the increase in time to perform the qudit interactions.  Generally, the ququart circuit time will approach the qutrit duration when the number of controls approaches a power of two.  Certain numbers of controls are not conducive to the tree-like structure, and will result in an excess number of gates.  When we examine the space time product, ququarts more efficiently utilizes the architecture than both qubits and qutrits.  In fact, the qubit to ququart space time product ratio is an average of 2.3 after the initial few increases in controls, which is inline with the expected increase of computational space from qubits to ququarts, $\text{log}_2(4) = 2$.  From qutrits to ququarts the expected increase would be $\text{log}_3(4) = 1.9$, and we find the ratio in space time product from qutrits to ququarts to be 1.43, an increase in the efficiency of the use of the device.

\subsection{Diminishing Benefits of Higher Levels}

When we move into higher level systems, specifically levels 5, 6 and 7, we do not see the benefits of using fewer qubits.  Since these generalized Toffoli gates have similar constructions to the ququart gate, and do not have the benefits of gate reduction, and do not see significant decreases in the number of gates or depth of the circuit.  As seen in Figure \ref{fig:pre-depth} and Figure \ref{fig:post-depth}, the ququint depth is the same as the ququart depth before and after insertion SWAPs for routing.  Interactions at the ququint level take longer than at the ququart level.  In Figure \ref{fig:circuit-duration} we see that the time for a ququint circuit to execute is generally greater than the time for a ququart circuit to execute.  This translates to an increased space time product when compared to ququarts.  The benefits afforded by the computation space affording by this level of qudit do not outweigh the cost of the complexity and operations required to reach this level over ququart circuits.

\label{sec:results}

\section{Conclusion}

We describe a strategy for qudit communication through efficient decomposition of SWAPs, and develop a straightforward compilation pipeline for qudit circuits, using realistic gate times for derived from optimal control.  Further, we were able to introduce a generalized Toffoli gate that use intermediate ququarts, ququints, following a structure that can be generalized to any number of qudit levels built on strategies developed for intermediate qutrits.  Finally, we compiled these circuits for potential architectures to evaluate the benefits at each level of qudit through examination in the change in gate count, depth, execution time, and space time product.

Increasing the highest possible level of a qudit offers opportunities to improve the utilization of the architecture.  From qubits to qutrits we see significant decreases in space time product to due reduction in ancilla use, even with increases in time of both interactions and SWAP gates.  This trend continues from qutrits to ququarts, where the extra computational space allows us to use two two qudit gates rather than a three qudit gate, reducing duration of the circuit, and the space time product of the circuit as well and improving the use of the architecture.  These advantages show that there are gains to be found by implementing these higher dimensional systems and integrating qudit compilation strategies to improve the usage of near term quantum devices.

\label{sec:conclusion}

\newcommand\acksname{Acknowledgments}
\specialcomment{acks}{%
  \begingroup
  \section*{\acksname}
  \phantomsection\addcontentsline{toc}{section}{\acksname}
}

\begin{acks}

Thank you to Max Siefert, Jason Chadwick and Natalia Nottingham for their helpful conversations in developing this work.
\end{acks}

\bibliographystyle{ieeetr}
\bibliography{ref}

\end{document}